\documentclass[12pt,a4paper,DIV12]{scrartcl}
\usepackage[T1]{fontenc}
\usepackage{lmodern}
\usepackage[british]{babel}
\usepackage{amsmath}
\usepackage{amssymb}
\usepackage{amsfonts}
\usepackage{color}
\usepackage{float}
\usepackage{hyperref}

\newcommand{\I}{\ensuremath{\mathrm{i}\hspace{1pt}}}
\newcommand{\Tr}[1]{\ensuremath{\operatorname{Tr}\left(#1\right)}}
\newcommand{\1}{\ensuremath{\boldsymbol{1}}}
\newcommand{\arxiv}[2]{[arXiv:\,\href{http://arxiv.org/abs/#1}{\texttt{#1}} [\texttt{#2}]]}
\newcommand{\arxivold}[1]{[arXiv:\,\href{http://arxiv.org/abs/#1}{\texttt{#1}}\,]}
\begin{document}
\title{Chiral perturbation theory for three-flavour lattice QCD
with isospin splitting
\vspace*{3mm}}
\author{S.~Engelnkemper and G.~M\"unster\\
\textit{\large Universit\"at M\"unster, Institut f\"ur Theoretische Physik}\\
\textit{\large Wilhelm-Klemm-Str.~9, D-48149 M\"unster, Germany}\\
\textit{\large E-Mail: engelnkemper@uni-muenster.de, munsteg@uni-muenster.de}
\vspace*{5mm}}

\date{March 8, 2016}

\maketitle

\begin{abstract}
\noindent
{\sffamily\bfseries\Large Abstract}\\[6pt]
An important tool for the analysis of results of numerical simulations of
lattice QCD is chiral perturbation theory. In Wilson chiral perturbation
theory the effects of the finite lattice spacing $a$ are taken into account.
In recent years the effects of isospin splitting on the masses of hadrons
have been investigated in Monte Carlo simulations. Correspondingly, in this
article we derive the expansions of the masses of the pseudoscalar mesons in
chiral perturbation theory at next-to-leading order for twisted mass lattice
QCD with three light quark flavours, taking the mass difference between the
up and down quarks into account. The results include terms up to orders
$m_q^2$ in the quark masses, $\Delta m^2$ in the mass splitting between up-
and down quarks, and $a^2$ in the lattice spacing, respectively.
\\[10mm]
\end{abstract}
\vspace{5mm}

\section{Introduction}

For investigations of QCD by means of numerical simulations on a lattice,
chiral perturbation theory has become an indispensable tool. It allows to
describe the dependence of hadron masses and matrix elements on the quark
masses in terms of expansions in powers of the quark masses, modified by
logarithmic terms. On the one hand this offers the possibility to
extrapolate the results of numerical simulations with larger quark masses
into the physical regime with small up- and down-quark masses. On the other
hand, comparing the quark mass dependence observed in Monte Carlo
calculations with the formulae from chiral perturbation theory yields the
values of fundamental low-energy constants of QCD, the Gasser-Leutwyler
coefficients. For reviews see \cite{Sharpe:2006pu,Golterman:2009kw}.

The discretisation of space-time on a lattice with lattice spacing $a$
implies a breaking of chiral symmetry and leads to discretisation effects in
observable quantities. In the case of lattice QCD with Wilson fermions the
latter are generally of order $a$. A way to reduce lattice
artifacts is to employ the formulation of twisted mass lattice QCD, which
implements a chirally rotated mass term
\cite{Frezzotti:1999vv,Frezzotti:2000nk}. Setting the twist angle to the value
$\pi/2$ (``full twist'') implies automatic $\mathcal{O}(a)$
Symanzik-improvement \cite{Frezzotti:2003ni,Frezzotti:2003xj}. The ETM
collaboration has done simulations with two mass-degenerate quark flavours,
representing the up and down quarks, and with $2+1+1$ quark flavours,
meaning a degenerate doublet of up and down quarks and a non-degenerate
doublet of charm and strange quarks, with full twist in both sectors
\cite{Shindler:2007vp,Baron:2010bv,Ottnad:2012fv}.

Chiral perturbation theory for twisted mass lattice QCD has been developed
in \cite{Munster:2003ba,Munster:2004am,Scorzato:2004da,Sharpe:2004ny} for
$N_{\textrm{f}} = 2$ quark flavours. It has been extended to $N_{\textrm{f}}
= 1+1$ non-degenerate quarks and $N_{\textrm{f}} = 2+1$, including the
strange quark, in \cite{Munster:2006yr}, and to $N_{\textrm{f}} = 2+1+1$,
with a degenerate doublet of up and down quarks, in \cite{Munster:2011gh},
and in \cite{Bar:2014bda} in the form of a charmless chiral Lagrangian.
The resulting formulae for meson masses reveal the dependence on the lattice
spacing $a$ in the form of additional terms proportional to powers of $a$.
Chiral perturbation theory taking lattice effects for Wilson fermions into
account is called Wilson chiral perturbation theory.

In recent years the isospin splitting produced by the small mass difference
between up and down quarks has come into the focus of numerical
investigations of lattice QCD
\cite{Blum:2010ym,Aoki:2012st,Borsanyi:2014jba}. This splitting contributes
to the mass splittings between charged and neutral hadrons. An additional
source for these mass splittings are electromagnetic effects, which have
been studied in the framework of lattice QCD + QED in 
\cite{Aoki:2012st,Ishikawa:2012ix,Horsley:2013qka,Borsanyi:2014jba}.

The purpose of this paper is to take the mass splitting between the light
quark masses into account in Wilson chiral perturbation theory for twisted
mass lattice QCD with $N_{\textrm{f}} = 1+1+1$ quark flavours. We calculate
the masses of the members of the pseudoscalar meson octet to next-to-leading
order (NLO). The small isospin mass splitting $\Delta m = m_{\text{d}} -
m_{\text{u}}$ is included to order $\Delta m$, and lattice effects are
included to order $a^2$.

The purpose of this article is twofold. On the one hand, the results include
the limit of a vanishing twist angle, which is ordinary untwisted lattice
QCD with three non-degenerate Wilson quarks. For this case our calculations
reveal the effects of the lattice discretisation on the isospin splitting
for pseudoscalar meson masses. In addition to that, even the further limit
of vanishing lattice spacing $a$ is of interest, as it includes the effects
of mixing in next-to-leading order, which are not completely included in the
original work \cite{Gasser:1985}.

On the other hand, simulations of twisted mass lattice QCD including strange
quarks, are in practice done by adding a fourth heavy charm quark and
implementing full twist in both the up-down and the strange-charm sector in
order to ensure automatic $\mathcal{O}(a)$ improvement in all observables.
For the case of the masses of pions, which are made of up quarks and down
quarks only, it turns out that full twist in the up-down sector alone is not
sufficient for $\mathcal{O}(a)$ improvement, and that the fully twisted
strange-charm sector is necessary. Our calculations make the influence of
unaccompanied strange quarks on the pion masses explicit and in this way
points to the need of including the charm quarks in the simulations.

\section{Chiral perturbation theory}

In twisted mass lattice QCD the twist in the quark sector has to be
implemented orthogonally to the mass splitting in order to maintain positivity
of the fermion determinant \cite{Frezzotti:2003xj,Frezzotti:2004pc}. In this
letter we introduce the mass twist in the u-d-sector via the third
Gell-Mann-matrix $\lambda_3$. Consequently the mass splitting is implemented
by means of $\lambda_1$. The strange quark is introduced as a standard Wilson fermion.
The three-flavour quark mass matrix is thus given
by
\begin{align}
\label{M-untwisted}
M &= \hat{m} \1  + (m_{\text{s}}-\hat{m}) \left(\frac{1}{3}\1 
- \frac{1}{\sqrt{3}}\lambda_8 \right) +  \frac{1}{2} \Delta m\lambda_1 
= \begin{pmatrix} \hat{m} & \frac{1}{2} \Delta m & 0 \\
  \frac{1}{2} \Delta m & \hat{m} & 0 \\
  0 & 0 & m_{\text{s}}
  \end{pmatrix}, \\
\hat{m} &\doteq \frac{m_{\text{u}} + m_{\text{d}}}{2}\,, \qquad 
\Delta m \doteq m_{\text{d}} - m_{\text{u}}.
\end{align}
The chiral rotation entailing the twisted mass is given by
\begin{align}
\label{M-twisted}
M\ &\rightarrow\ \exp\left[\I\:\!\gamma_5 \lambda_3 \frac{\omega}{2} \right]
M \exp\left[\I\:\!\gamma_5 \lambda_3 \frac{\omega}{2} \right] 
= \begin{pmatrix} \tilde m + \I\:\!\gamma_5 \mu & \frac{1}{2} \Delta m & 0 \\
\frac{1}{2}\Delta m & \tilde m - \I\gamma_5\:\!\mu & 0 \\
0 & 0 & m_{\text{s}}  \end{pmatrix},\\
\mu &= \hat{m}\,\sin(\omega)\,, \qquad 
\tilde m = \hat{m}\cos(\omega)\,, \qquad 
\omega = \text{arctan}\left( \frac{\mu}{\tilde m} \right)\,.
\end{align}

The effective Lagrangian describing the physics of the pseudoscalar mesons
is constructed via the power-counting scheme of \cite{Bar:2003mh},
accounting for terms of order $\{m,a\}$ in leading order (LO), and
$\{m^2,ma,a^2\}$ in next-to-leading order (NLO), respectively. The
Lagrangian is formulated in terms of the matrix-valued field
\begin{equation}
U \doteq \exp\left[\frac{\I}{F_0}\sum_{a=1}^8 \lambda_a \phi_a \right]\,.
\end{equation}
It contains the meson fields $\phi_a$ and the low-energy constant (LEC)
$F_0$, which equals the pion decay constant to leading order. The quark
masses and the lattice spacing $a$ enter the Lagrangian through the
variables
\begin{align}
\chi \doteq 2 B_0 M\,, 
\qquad \rho \doteq \rho_0 \1 = 2 W_0 a \1\,,
\end{align}
where $B_0$ and $W_0$ are additional LECs. To leading order the chiral
effective Lagrangian reads
\begin{equation}
\mathcal{L}_{\text{LO}} 
= \frac{F_0^2}{4} \Tr{\partial_{\mu} U \, \partial_{\mu} U^\dagger }
- \frac{F_0^2}{4} \Tr{\chi U^\dagger + U \chi^\dagger}
- \frac{F_0^2}{4} \Tr{\rho \, U^\dagger + U \rho^\dagger}.
\end{equation}

In numerical simulations of twisted mass lattice QCD a chirally rotated mass
matrix like Eq.~(\ref{M-twisted}) has to be used. On the other hand, in
chiral perturbation theory it is more convenient to employ the
\textit{physical basis}. It is obtained by applying the inverse of the above
chiral rotation, such that the mass matrix retains its original form
Eq.~(\ref{M-untwisted}), and the lattice term goes over to
\begin{align}
\rho (\omega) &= \exp\left[\I \lambda_3 \frac{\omega}{2} \right] \rho
\exp\left[\I \lambda_3 \frac{\omega}{2} \right]
= \begin{pmatrix} 
\tilde \rho + \I\:\!\rho_3 & 0 & 0 \\
0&  \tilde \rho - \I\:\!\rho_3 & 0 \\
0 & 0 & \rho_0  
\end{pmatrix}\nonumber\\
\label{rho}
&= \frac{1}{3} \left(2\tilde\rho + \rho_0 \right) \1 
- \frac{1}{\sqrt{3}} \left(\rho_0 - \tilde\rho \right)\lambda_8 
+ \I \rho_3\lambda_3\,,\\
 \tilde\rho &= \rho_0 \cos(\omega)\,,  \qquad
\rho_3 = \rho_0\sin(\omega)\,.
\end{align}

In order to set up chiral perturbation theory for the computation of meson
masses and other observables, the effective Lagrangian has to be expanded
around its minimum. In contrast to the case of continuous space-time, where
the minimum is located at vanishing meson fields, $\phi_a = 0$, its position
is shifted for twisted mass lattice QCD and is located at
\begin{align}
\check\phi_i = 0 \quad \text{for} \quad i \neq 3, \qquad 
\check\phi_3 = \frac{F_0\rho_3}{\hat{\chi} + \tilde\rho} \,,
\end{align}
and the corresponding value of the field $U$ is
\begin{align}
U_0 \doteq \exp\left[\frac{\I}{F_0} \lambda_3 \check\phi_3 \right].
\end{align}
It has turned out that a suitable parametrisation of the field $U$, taking
the expansion point $U_0$ into account, is given by
\begin{align}
\label{Uprime}
U = U_0^{1/2} U' U_0^{1/2},
\end{align}
such that the minimum of the Lagrangian is now at $U' = \1$. Expanding in
terms of the shifted meson fields
\begin{equation}
U' \doteq \exp\left[\frac{\I}{F_0}\sum_{a=1}^8 \lambda_a \phi'_a \right]
\end{equation}
with the help of the Baker-Campbell-Hausdorff series \cite{Reinsch:2000} has
the advantage that spurious three-point vertices are eliminated and that
just one quartic vertex remains in the Feynman rules, which simplifies the
loop calculations a lot.

In order to derive the meson masses in NLO, the effective Lagrangian has to
be extended by terms of the next higher chiral dimension. Neglecting
external fields and constant terms, this leads to the NLO lattice $\chi$PT
Lagrangian \cite{Bar:2003mh,Munster:2003ba}
\begin{align}
\label{L-NLO}
\mathcal{L}_{\text{NLO}} 
= & \quad \mathcal{L}_{\text{LO}} 
- L_1 \left[\Tr{\partial_{\mu} U \partial_{\mu} U^\dagger}\right]^2
- L_2 \left[\Tr{\partial_{\mu} U \partial_\nu U^\dagger}\right]^2
- L_3 \Tr{[\partial_{\mu} U \partial_{\mu} U^\dagger]^2} \nonumber\\
& + L_4 \Tr{\partial_{\mu} U \partial_{\mu} U^\dagger} \Tr{\chi^\dagger U 
  + U^\dagger\chi}
  + W_4 \Tr{\partial_{\mu} U \partial_{\mu} U^\dagger} \Tr{\rho^\dagger U 
  + U^\dagger \rho} \nonumber\\
& + L_5 \Tr{\partial_{\mu} U \partial_{\mu} U^\dagger (U \chi^\dagger 
  + \chi U^\dagger)}
  + W_5 \Tr{\partial_{\mu} U \partial_{\mu} U^\dagger (U \rho^\dagger 
  + \rho \, U^\dagger)} \nonumber\\
& - L_6 \left[\Tr{\chi^\dagger U + U^\dagger\chi}\right]^2
- W_6 \Tr{\chi^\dagger U + U^\dagger\chi} 
      \Tr{\rho^\dagger U + U^\dagger \rho} \nonumber\\
& - L_7 \left[\Tr{\chi^\dagger U - U^\dagger\chi}\right]^2
- W_7 \Tr{\chi^\dagger U - U^\dagger\chi} 
      \Tr{\rho^\dagger U - U^\dagger \rho} \nonumber\\
& - L_8 \Tr{\chi^\dagger U \chi^\dagger U + U^\dagger\chi U^\dagger\chi}
- W_8 \Tr{\rho^\dagger U\chi^\dagger U 
          + U^\dagger \rho \, U^\dagger\chi} \nonumber\\
& - W'_6 \left[\Tr{\rho^\dagger U + U^\dagger \rho}\right]^2
- W'_7 \left[\Tr{\rho^\dagger U - U^\dagger \rho}\right]^2 \nonumber\\
& - W'_8 \Tr{\rho^\dagger U \rho^\dagger U 
            + U^\dagger \rho \, U^\dagger \rho}.
\end{align}

The masses in NLO receive tree-level contributions from
$\mathcal{L}_{\text{NLO}}$ and one-loop contributions from
$\mathcal{L}_{\text{LO}}$. The loop contributions produce divergences that
require renormalisation of the low energy coefficients $L_i$, $W_i$ and
$W'_i$.

The necessary steps for deriving the meson masses are as follows. The mass
matrix (\ref{M-untwisted}) and the twisted lattice term (\ref{rho}) are
inserted in the Lagrangian (\ref{L-NLO}), and the resulting expression is
expanded in the meson fields $\phi_a$ up to linear terms in order to find
the minimum. The minimum of the Lagrangian at NLO is given by
\begin{equation}
\check\phi_3 =
\ F_0 \rho_3 \,\frac{F_0^2 + 8 W_6(2 \hat{\chi} + \chi_{\text{s}}) 
+ 8 W_8 \hat{\chi}}
{F_0^2\hat{\chi} + 16 L_6 (2 \hat{\chi} + \chi_{\text{s}}) \hat{\chi} 
 + 16 L_8 (\hat{\chi}^2 + \Delta\chi^2)} 
+ \mathcal{O}(a^2)\,.
\end{equation}
As this quantity enters the calculation only quadratically, its terms $\sim
a^2$ can be neglected. Next the parametrisation of the fields according to
Eq.~(\ref{Uprime}) is implemented, such that the minimum of the Lagrangian
occurs at the origin of the shifted meson fields. Expanding around this
minimum up to quadratic terms in the fields gives us the inverse meson
propagator at tree level. To LO the inverse propagator reads
\begin{equation}
\I G^{-1}(p^2) = p^2\1 + \mathfrak{M}_{\text{LO}} \,.
\end{equation} 
The LO mass squared matrix $\mathfrak{M}_{\text{LO}}$ contains non-diagonal
terms that generate mixings in the $\pi^0$-$\eta$ sector and among the kaon
fields. The loop calculations require the LO propagator and the quartic
vertices. The vertices are obtained by expanding the NLO Lagrangian up to
quartic terms in the shifted meson fields, which leads to complicated
expressions containing momentum dependent terms.

At NLO the inverse meson propagator receives additional tree-level
contributions and self-energy terms from the loop diagrams,
\begin{equation}
\I G^{-1}(p^2) = p^2\1 + \mathfrak{M}_{\text{LO}} - A + p^2 B\,.
\end{equation} 
The matrices $A$ and $B$ are rather complicated and contain various
off-diagonal mixing terms. The divergent contributions in the self energies
can be separated from the finite ones and absorbed into the NLO renormalised
LECs $L_i^{\text{r}}$ via a modified minimal subtraction scheme as shown in
\cite{Gasser:1985}. Field renormalisation can be achieved by writing the
inverse propagator as
\begin{equation}
\I G^{-1}(p^2) = \bigg(\1 + \frac{B}{2} \bigg)
\bigg[p^2\1 + \mathfrak{M}_{\text{LO}} - A  
- \frac{1}{2} \{B,\mathfrak{M}_{\text{LO}}\} \bigg]
\bigg(\1 + \frac{B}{2} \bigg)+ \mathcal{O}(p^6)\,,
\end{equation}
such that the expression in square brackets contains the normalised kinetic
term. The squared meson masses are then obtained by diagonalising the 
matrix $[\mathfrak{M}_{\text{LO}} - A - \frac{1}{2}
\{B,\mathfrak{M}_{\text{LO}}\}]$.

\section{Results for the isospin splittings}

To LO in the limit $\Delta m \rightarrow 0$ the method outlined above
reproduces the known results \cite{Munster:2006yr}
\begin{align}
\bar m_{\pi}^2 =
&\ 2 B_0 \hat{m} + 2 a W_0 \cos(\omega) 
 + \frac{a^2 W_0^2 \sin^2(\omega)}{B_0 \hat{m}}
 + \mathcal{O}(a^3)\\
\bar m_{K}^2 =
&\ B_0 (\hat{m} + m_{\text{s}}) + a W_0 (1 + \cos(\omega)) 
 + \frac{1}{2} \frac{a^2 W_0^2 \sin^2(\omega)}{B_0 \hat{m}}
 + \mathcal{O}(a^3)\\
\bar m_{\eta}^2 =
&\ \frac{1}{3} \left[2 B_0 (\hat{m} + 2m_{\text{s}}) 
 + 2 a W_0 (2 + \cos(\omega)) 
 + \frac{a^2 W_0^2 \sin^2(\omega)}{B_0 \hat{m}} \right]
 + \mathcal{O}(a^3)
\end{align}
We list these expressions here since they are used as convenient
abbreviations in the NLO results. 

With our choice of the generators for the quark mass splitting and twist,
the neutral pion is $\pi_1 \equiv \pi^0$, and the charged pions $\pi^{\pm}$
are linear combinations of the mass eigenstates $\pi_2$ and $\pi_3$. The
charged pion $\pi_2$ represents the component that does not mix with the
$\eta$ and is ``orthogonal'' to twist ($\lambda_3$) and to mass splitting
($\lambda_1$), and is therefore least affected by these factors.
The mass of the charged pion $\pi_2$ is given in NLO by

\begin{align}
\label{pion2mass}
M_{\pi_2}^2
&= \bar m_{\pi}^2
+ \frac{32 B_0^2}{F_0^2} 
\Big[ \big( -2 L_4^{\text{r}} - L_5^{\text{r}} + 4 L_6^{\text{r}}
+2 L_8^{\text{r}} \big) \hat{m}^2
+ \big(- L_4^{\text{r}} + 2 L_6^{\text{r}} \big) 
   \hat{m} \, m_{\text{s}} \Big] \nonumber\\
&+ \frac{32 a B_0 W_0}{F_0^2} 
   \Big[ \big(-W_4^{\text{r}} + W_6^{\text{r}} \big) \hat{m} 
   + \big(- L_4^{\text{r}} + W_6^{\text{r}} \big)
            m_{\text{s}} \cos(\omega) \nonumber\\
&\qquad\qquad\qquad \, + \big(-2 L_4^{\text{r}} - L_5^{\text{r}}
 -2 W_4^{\text{r}} - W_5^{\text{r}} +4 W_6^{\text{r}} + 2 W_8^{\text{r}} \big)
 \hat{m} \cos(\omega) \Big]\nonumber\\
&+ \frac{32 a^2 W_0^2}{F_0^2} 
\bigg[-2 W_4^{\text{r}} - W_5^{\text{r}} + 4 {W_6'}^{\text{r}} 
 + 2 {W_8'}^{\text{r}}
 + \big(- W_4^{\text{r}} + 2 {W_6'}^{\text{r}} \big) \cos(\omega)\nonumber\\
&\qquad\qquad\quad \ \, 
 + (-4 L_6^{\text{r}} - 2 L_8^{\text{r}}
       + 4 W_6^{\text{r}} + 2 W_8^{\text{r}}
       - 4 {W_6'}^{\text{r}} - 2 {W_8'}^{\text{r}}) \sin^2(\omega)\nonumber\\
&\qquad\qquad\quad \ \, +\big(-L_4^{\text{r}} - 2 L_6^{\text{r}}
        +2 W_6^{\text{r}}\big)
         \left[\frac{m_{\text{s}}}{2 \hat{m}}\right]
            \sin^2(\omega) 
\bigg]\nonumber\\
&+ \frac{\bar m_{\pi}^2}{32\pi^2 F_0^2}
   \bigg[\bar m_{\pi}^2 \ln\bigg(\frac{\bar m_{\pi}^2}{\Lambda^2}\bigg)
   - \frac{\bar m_{\eta}^2}{3} 
     \ln\bigg(\frac{\bar m_{\eta}^2}{\Lambda^2} \bigg) \bigg] 
+ \mathcal{O}(\Delta m^3,a^3).
\end{align}

In NLO the pions appear to lose their automatic
$\mathcal{O}(a)$-improvement at full-twist, due to the presence of the term
proportional to $(-W_4^{\text{r}} + W_6^{\text{r}}) a$. This term stems from
the presence of the strange quark and is independent of $\omega$.

We also note that in the pion masses dependencies on the strange quark mass
$m_{\text{s}}$ show up, that are independent of $\omega$ and persist in the
continuum limit. As the pions are made from up and down quarks, this term
proportional to $m_{\text{s}}$ is counterintuitive. In the context of QCD in
the continuum it has its origin in the fermion determinant or, in terms of
Feynman diagrams, in strange quark loops, that lead to NLO terms in the
effective Lagrangian. The associated LECs $L_4^{\text{r}}$ 
and $L_6^{\text{r}}$ have very small empirical values,
however \cite{Bijnens:2014lea}.

In the calculation of the NLO pion masses for non-degenerate quark masses we
find that the mass splitting between the neutral pion and the charged pion
$\pi_2$ is induced by the $\pi^0$-$\eta$ mixing, and surprisingly appears to
be at least of order $\Delta m^2$. These terms are manageable
in the tree-level contributions, but lead to quite complicated dependences
in the loop contributions, in particular in the $\mathcal{O}(a^2)$ terms.
Since the quark mass difference $\Delta m^2$ is very small, we decide to list
the pion mass splittings including the $\Delta m^2$ terms up to order
$a$ only, neglecting $\mathcal{O}(a^2)$ terms. Then the mass
difference between the neutral pion $\pi_1$ and the charged pion $\pi_2$ is

\begin{align}
M_{\pi_1}^2&- M_{\pi_2}^2 = B_0 \Delta m^2 \bigg\{
\frac{1}{4 (m_{\text{s}} - \hat{m})}
\left[1 -
\frac{a W_0 (1 - \cos \omega)}{B_0 (m_{\text{s}} - \hat{m})}\right]
\nonumber\\
&+ \frac{4 B_0}{3 F_0^2}
(- 3 L_4^{\text{r}} + 2 L_5^{\text{r}} + 6 L_6^{\text{r}} 
 - 78 L_7^{\text{r}} - 18 L_8^{\text{r}})
+ \frac{4 B_0}{F_0^2}
\left(\frac{\hat{m}}{m_{\text{s}} - \hat{m}} \right)
(- 3 L_4^{\text{r}} - 2 L_5^{\text{r}} + 6 L_6^{\text{r}} )\nonumber\\
&+ \frac{4 a W_0}{F_0^2 (m_{\text{s}} - \hat{m})} \big[
(- L_5^{\text{r}} - 3 W_4^{\text{r}} - W_5^{\text{r}} + 3 W_6^{\text{r}}) 
\nonumber\\
&\qquad + (L_4^{\text{r}} + L_5^{\text{r}} + 20 L_7^{\text{r}} 
+ 4 L_8^{\text{r}} + 2 W_4^{\text{r}} + W_5^{\text{r}} - 3 W_6^{\text{r}} 
- 10 W_7^{\text{r}} - 2 W_8^{\text{r}} )
(1 - \cos \omega) \nonumber\\
&\qquad + \left(\frac{\hat{m}}{m_{\text{s}} - \hat{m}} \right)
(3 L_4^{\text{r}} + L_5^{\text{r}} + 4 L_8^{\text{r}} + W_5^{\text{r}} 
- 3 W_6^{\text{r}} - 2 W_8^{\text{r}} )
(1 - \cos \omega) \big] \nonumber\\
&- \frac{B_0}{64 \pi^2 F_0^2}
\bigg[ \ln \bigg(\frac{\bar{m}_{K}^2}{\Lambda^2}\bigg) + 1 \bigg]
\nonumber\\
&- \frac{1}{192 \pi^2 F_0^2 (m_\text{s} - \hat{m})}
\left[1 -
\frac{a W_0(1 - \cos \omega)}{B_0 (m_{\text{s}} - \hat{m})}\right] \nonumber\\
&\qquad
\bigg[\bar{m}_{\pi}^2 \ln \bigg(\frac{\bar m_{\pi}^2}{\Lambda^2}\bigg) 
- 3\bar{m}_{\eta}^2 \ln \bigg(\frac{\bar m_{\eta}^2}{\Lambda^2} \bigg) 
- \frac{1}{2} (3 \bar{m}_{\eta}^2 - \bar{m}_{\pi}^2) 
\ln \bigg(\frac{\bar{m}_{K}^2}{\Lambda^2}\bigg)
+ \frac{3}{2} (\bar{m}_{\pi}^2 + \bar{m}_{\eta}^2) \bigg] \bigg\}
\nonumber\\
&+ \mathcal{O}(\Delta m^3,a^2).
\end{align}
We observe that at full twist, $\cos \omega = 0$, also terms of order $a$
are present. They are not proportional to $m_\text{s}$. The presence of these
terms is a consequence of the fact that the strange quark is not a member of
a twisted doublet. Namely, if the strange quark would be accompanied by a 
fourth quark, forming together a fully twisted doublet, automatic 
$\mathcal{O}(a)$ improvement would forbid the existence of such terms.

Due to the mass-twist the charged pions represented by the fields $\pi_2$
and $\pi_3$ also gain different masses, which leads to the splitting
\begin{equation}
M_{\pi_3}^2 - M_{\pi_{2}}^2 
= \frac{32 a^2 W_0^2}{F_0^2}
\left( -4 L_6^{\text{r}} - 2 L_8^{\text{r}} + 4 W_6^{\text{r}}
+ 2 W_8^{\text{r}} - 4 {W_6'}^{\text{r}} - 2 {W_8'}^{\text{r}}
\right) \sin^2(\omega)
+ \mathcal{O}(a^3).
\end{equation}
This mass difference is a pure lattice artifact that vanishes in the
continuum limit $a \rightarrow 0$.

We have also calculated the mass splitting between the kaon masses in NLO
including terms of order $a^2$. As the characteristic features show up at
order $a$ already, we display the mass splitting between neutral kaons $K_6,
K_7$ and charged kaons $K_4, K_5$, including terms linear in $a$, which is

\begin{align}
M^2_{K^0}&- M^2_{K^{\pm}} = 
B_0 \Delta m \bigg\{ 1 \nonumber\\
& + \frac{16 B_0}{F_0^2} 
  \bigg[ \big(- 2 L_4^{\text{r}} - L_5^{\text{r}} + 4 L_6^{\text{r}} 
              + 2 L_8^{\text{r}} \big) \hat{m} 
+ \big( - L_4^{\text{r}} - L_5^{\text{r}}
                       + 2 L_6^{\text{r}} + 2 L_8^{\text{r}} \big) 
                           m_{\text{s}} \bigg]  \nonumber\\
& + \frac{8 a W_0}{F_0^2} 
  \bigg[ \big( - 2 L_5^{\text{r}} - 6 W_4^{\text{r}} - 2 W_5^{\text{r}} 
               + 6 W_6^{\text{r}} + 4 W_8^{\text{r}} \big) \nonumber\\
& \hspace{4em} + \big( L_5^{\text{r}} + 4 W_4^{\text{r}}
  + W_5^{\text{r}} - 4 W_6^{\text{r}} - 2 W_8^{\text{r}} \big)
(1 - \cos \omega) \bigg] \nonumber\\
& + \frac{1}{32 \pi^2 F_0^2} \bigg[ 
\frac{2 \bar{m}_{\eta}^2}{3} 
   \ln\bigg( \frac{\bar{m}_{\eta}^2}{\Lambda^2} \bigg)
\nonumber\\
&\qquad + \frac{1}{B_0 (m_{\text{s}} - \hat{m})} 
\frac{\bar{m}_K^2}{2} 
\bigg[ \bar{m}_{\eta}^2 \ln\bigg( \frac{\bar{m}_{\eta}^2}{\Lambda^2} \bigg)
-\bar{m}_{\pi}^2 \ln\bigg( \frac{\bar{m}_{\pi}^2}{\Lambda^2} \bigg)
\bigg]
\bigg( 1 - \frac{a W_0 (1 - \cos \omega )}
  {B_0 (m_{\text{s}} - \hat{m})} \bigg)
\bigg] \bigg\} \nonumber\\
& + \mathcal{O}(\Delta m^2,a^2).
\end{align}
It vanishes in the limit $\Delta m = 0$, as it should. In the continuum
limit the expression reduces to that given in \cite{Gasser:1985}.

In contrast to the case of pions, the kaon mass splitting has contributions
linear in $\Delta m$, part of which is produced by pion and eta loops.
It is thus significantly more sensitive to the tiny quark mass difference
$\Delta m$ than the pion mass splitting. Therefore the kaon masses represent
observables ideally suited for numerical investigations of the quark mass
difference $\Delta m$ in lattice QCD. For studies with three flavours of
Wilson quarks the formula shows the lattice artifacts proportional
to $a \Delta m$.

We have also obtained the expression for the mass of the $\eta$ meson in the
case of non-degenerate quark masses. It is quite lengthy, and as it does not
reveal any new characteristic features compared to the pion masses, we
refrain from presenting it here.

\section{Conclusions}

To summarise, the masses of the pseudoscalar mesons for the case of
non-degenerate up, down and strange quarks have been obtained in chiral
perturbation theory for twisted mass lattice QCD including lattice terms
quadratic in the lattice spacing $a$ and in $\Delta m$. Automatic
$\mathcal{O}(a)$ improvement of the pion masses at maximal twist in the
up-down sector is lost. At maximal twist, terms of order $a$ are present,
which have two different origins. On the one hand, at NLO the pion masses
get an equal contribution of order $a$ that stems from the presence of the
strange quark in the effective Lagrangian.
On the other hand, the inverse propagator contains off-diagonal
terms that represent a mixing between the neutral pion and the $\eta$ meson.
These yield a contribution to the neutral pion mass which is proportional
to $\Delta m^2$. Apart from continuum terms ($a^0$) it has terms of order $a$
that spoil automatic $\mathcal{O}(a)$ improvement at maximal twist.
We present the mass splitting between neutral and charged pions
including terms of orders $\Delta m^2 $ and $a$. In addition, the
twisted mass generates a mass difference between the charged pions of order
$a^2$. 

Finally, the lattice corrections to the mass splitting between
neutral and charged kaons, which are proportional to $\Delta m$, are
presented including terms of order $a$. In the kaon sector there is also
mixing, in this case between $K_4$ and $K_6$ and between $K_5$ and $K_7$.
As for the pions, it leads to terms of order $a$ that spoil automatic
$\mathcal{O}(a)$ improvement.

If the strange quark were introduced with a partner (charm quark),
associated with another twist in this heavy sector, the lattice
artefacts of order $a$ and higher would look quite different. In
particular, full twist in both sectors would ensure absence of all
$\mathcal{O}(a)$ terms. Our results make the idea that such
inclusion of a twisted fourth quark is advantageous for reducing
unwanted lattice artifacts explicit.


%
\end{document}